\documentclass{PoS}

\usepackage{longtable}

\newcommand*{\no}{\noindent}
\newcommand*{\bea}{\begin{eqnarray}}
\newcommand*{\eea}{\end{eqnarray}}
\newcommand*{\be}{\begin{equation}}
\newcommand*{\ee}{\end{equation}}
\newcommand*{\pd}{\partial}
\newcommand*{\pdm}{\pd_{\mu}}

\newcommand*{\nn}{\nonumber}

\title{Scalar-matter--gluon interaction}

\ShortTitle{Scalar-matter--gluon interaction}

\author{\speaker{Axel Maas}\thanks{Supported by the FWF under grant number M1099-N16 and the DFG under grant number MA 3935/5-1.}\\
        Institute for Theoretical Physics, Friedrich-Schiller-University Jena, Max-Wien-Platz 1, D-07743 Jena, Germany\\
        E-mail: \email{axelmaas@web.de}}

\abstract{A full non-perturbative treatment of gauge theories requires to include matter fields on equal footing with the gauge fields. Scalar matter can act as a role model for generic matter, as many questions, e.\ g.\ confinement, can be posed without referring to a particular Lorentz structure. Due to their rather simple structure they are also useful to develop methods.

One possible way to describe gauge theories beyond perturbation theory is based on correlation functions. After a short discussion of the setup, lattice gauge theory is used to analyze the interaction of gluons with quenched fundamental and adjoint scalars. Both the two-point and three-point correlation functions for massive and massless adjoint and fundamental scalars will be determined, in minimal Landau gauge. The findings are in agreement with the possibility that scalars are only slightly affected by the interaction with gluons. The results are compared briefly with dynamical, massive scalars, showing no significant changes in the confinement region compared to the quenched case.}

\FullConference{The many faces of QCD\\
		November 2-5, 2010\\
		Gent Belgium}

\begin{document}

\section{Introduction}

A full non-perturbative description of gauge theories, with QCD as a prime example, at all energy scales is arguably one of the central goals of field theory. One possibility to achieve this goal is a combination of various non-perturbative methods, in particular lattice calculations and functional methods, to obtain systematic control over this very difficult problem. The basic elements in such an approach are the (gauge-dependent) correlation functions of the elementary degrees of freedom, which are then combined to give access to measurable observables \cite{Fischer:2006ub,Maas:2010gi}.

Of course, such an approach requires particularly good control over the most simple correlation functions, the two-point functions and primitively divergent vertices. For fermions, this has turned out to be quite a formidable task \cite{Fischer:2006ub,Alkofer:2008tt,Kamleh:2007ud,Aguilar:2010ad,Kizilersu:2006et,Zwanziger:2010iz}. It appears therefore useful to also investigate the more simple case of scalar particles to improve control over the methods \cite{Fister:2010yw,Alkofer:2010tq,Maas:2010nc}. In addition, scalars are interesting entities on their own, since they permit to study confinement \cite{Caudy:2007sf,Maas:2010nc,Fister:2010yw,Philipsen:1998de}, the Higgs effect \cite{Caudy:2007sf,Maas:2010nc}, and triviality of gauge field theories \cite{Callaway:1988ya}, among other important problems \cite{Caudy:2007sf,Maas:2010nc}.

The basic question in this endeavor is what the properties of the elementary correlation functions are. To provide a first investigation and supply truncation schemes for functional equations with valuable input it is thus worthwhile to investigate them in the limited setting of lattice calculations. Here, preliminary results will be presented for the quenched case. The investigation of the quenched case is not motivated by the computation time as in the case of fermions, but rather it is physically interesting. It is known that quenched adjoint and fundamental charges show quite a different behavior in connection to confinement and screening\cite{Greensite:2003bk}: Fundamental charges are connected by a linearly rising Wilson line, while the one between adjoint charges is asymptotically flat. It has been argued that this could manifest itself in the two-scalar-gluon vertex in the quenched case \cite{Fister:2010yw}. Furthermore, it is possible that whether a particle belongs to the physical Hilbert space can be read off from the properties of the corresponding propagators \cite{Alkofer:2003jj,Maas:2004se}. Therefore, the propagators are also quite interesting to investigate. Finally, in general it could be expected that there could be a difference between massive and massless scalars \cite{Maas:2005rf,Cucchieri:2001tw}, and therefore investigating both cases is worthwhile.

In the following, preliminary results will be presented for these quantities. Details of the setup are given in section \ref{sec:setup}, and the results will be presented in section \ref{sec:quenched}. These will be compared to existing results \cite{Maas:2010nc,Cucchieri:2001tw} for the unquenched case in section \ref{sec:unquenched}. A short summary in section \ref{sec:sum} will complete this presentation.

\section{Setup}\label{sec:setup}

The simulation of quenched scalars can essentially be performed like the one of quenched fermions \cite{Gattringer:2010zz}. The only input necessary are Yang-Mills background fields, which will be generated using the techniques described in \cite{Cucchieri:2006tf}. The calculations will be performed for two, three, and four dimensions.

The quenched propagator is then given by \cite{Maas:2010nc,Maas:unpublished}
\be
D^{ijR}_S(p)=<D^{ij-1}>(p)\nn,
\ee
\no where $R$ denotes the representation and $D^{ij}$ is the covariant Laplacian, symbolically in the continuum
\be
D^{ij}=(\delta^{ik}\pdm+g\tau^{ik}_a A_\mu^a)(\delta^{jk}\pdm+g\tau^{jk}_b A_\mu^b)+m^2\delta^{ij}
\ee
\no with $i$ and $j$ being either fundamental or adjoint indices. See \cite{Maas:2010nc,Maas:unpublished} for the precise lattice implementation. The bare mass $m$ will be selected to be 0, 100, 1000, or 10000 MeV. The (non-amputated) two-scalar-gluon vertices can be constructed in analogy to the ghost-gluon vertex \cite{Cucchieri:2004sq} and read \cite{Maas:unpublished}
\be
\Gamma_\mu^{aij}(p,q,p+q)=<A_\mu^a(p)D^{ij-1}(q,p+q)>\nn.
\ee
\no In the present case the gauge group chosen is SU(2). Thus, the color structure can only be the tree-level one $f^{abc}$. Furthermore, as in the case of the ghost-gluon vertex \cite{Cucchieri:2006tf}, the unamputated vertex has only one tensor structure, which will be denoted by $A$, and is one at tree-level. As for operators involving ghost fields, a conjugate gradient algorithm can be used to determine the inverted operators in momentum space \cite{Cucchieri:2006tf}. Note that due to the absence of trivial zero modes, the inverse covariant Laplacian could in principle be evaluated at zero momentum. However, due to the influence of constant modes, which are sensitive to the boundary conditions, this will not be done here. Note further that the algorithm is also working when applied to the zero mass case, and the inversion is numerically cheaper the larger the bare mass.

Of course, all of these quantities are zero if the background field is not gauge-fixed. The gauge chosen here is minimal Landau gauge, as this gauge is numerically the cheapest one. It should be noted that this gauge is strictly speaking not the one used for the calculations using functional methods in \cite{Fister:2010yw,Alkofer:2010tq}. See \cite{Fischer:2008uz,Maas:2009se,Maas:2010wb,Cucchieri:2007rg,Cucchieri:2008fc,Bogolubsky:2009dc,Binosi:2009qm,Bornyakov:2008yx,Boucaud:2008ji,Dudal:2008sp} for a detailed discussion of other options. However, at the volumes employed here no significant influence of the gauge choice is expected. Furthermore, the scalar fields do not seem to be very sensitive to such gauge choices compared to the gauge fields \cite{Maas:2010nc}. In addition, the volumes in this preliminary report are too small yet for being able to probe the very far infrared. The implementation of minimal Landau gauge used here is described in \cite{Cucchieri:2006tf}.

This is sufficient to determine the non-renormalized correlation functions. For the renormalization of the propagator both a wave-function and a mass renormalization are necessary. This will be performed as described in \cite{Maas:2010nc}, demanding that at $\mu=2$ GeV the propagator and its first derivative coincide with the tree-level propagator with the tree-level mass. In contrast to the ghost-gluon vertex, the two-scalar-gluon vertex has to be renormalized as well. Usually, a symmetric point or the Thomson-limit are useful choices for such a renormalization. Both of them are not readily accessible in lattice calculations when performing calculations from two to four dimensions \cite{Maas:2007uv}, as will be done here. Thus, instead the more easily accessible \cite{Cucchieri:2006tf} point $pq=0$ and $p^2=q^2=1.5$ GeV will be used, where the vertex is required to be at tree-level. It turns out that the vertex is sensitive to cut-off effects for a lattice spacing $a^{-1}$ below approximately 2 GeV \cite{Maas:unpublished}. Thus the renormalization at these lattice spacings can produce additional cut-off effects, which should be kept in mind.

This completes the setup for the calculations. More details can be found in \cite{Maas:2010nc,Maas:unpublished}.

\section{Quenched scalars}\label{sec:quenched}

\subsection{Propagators}

\begin{figure}
\includegraphics[width=0.5\linewidth]{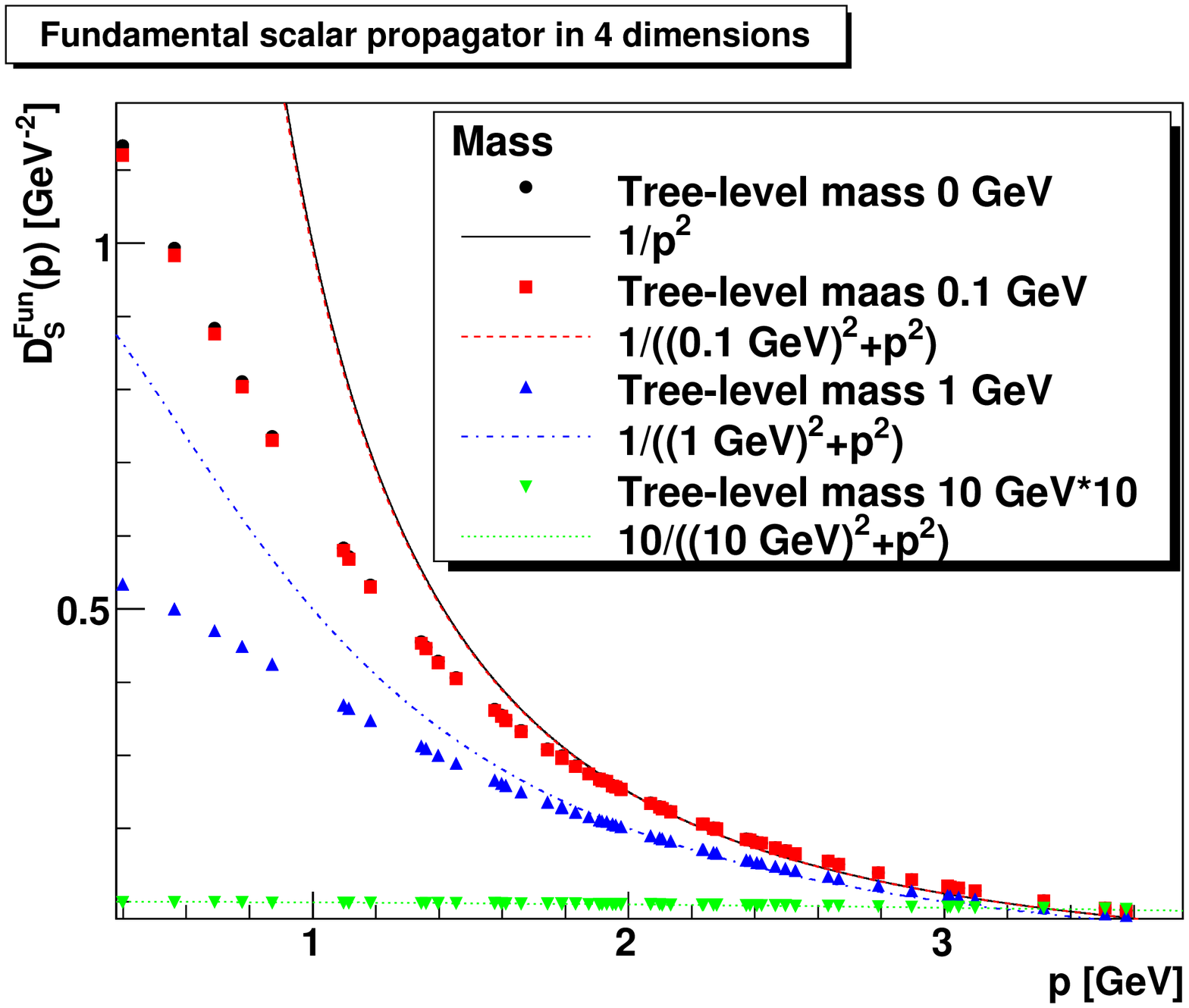}\includegraphics[width=0.5\linewidth]{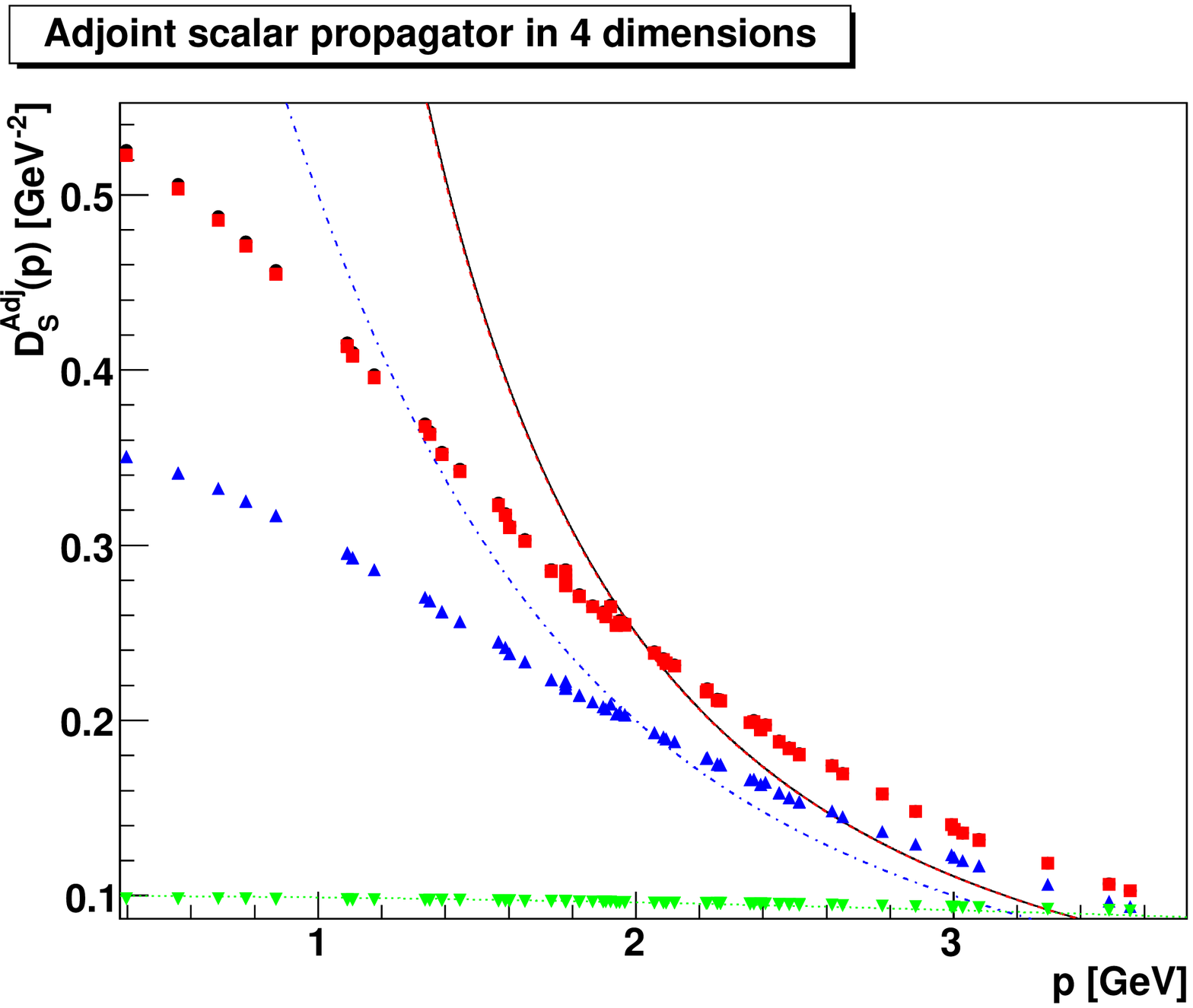}\\
\includegraphics[width=0.5\linewidth]{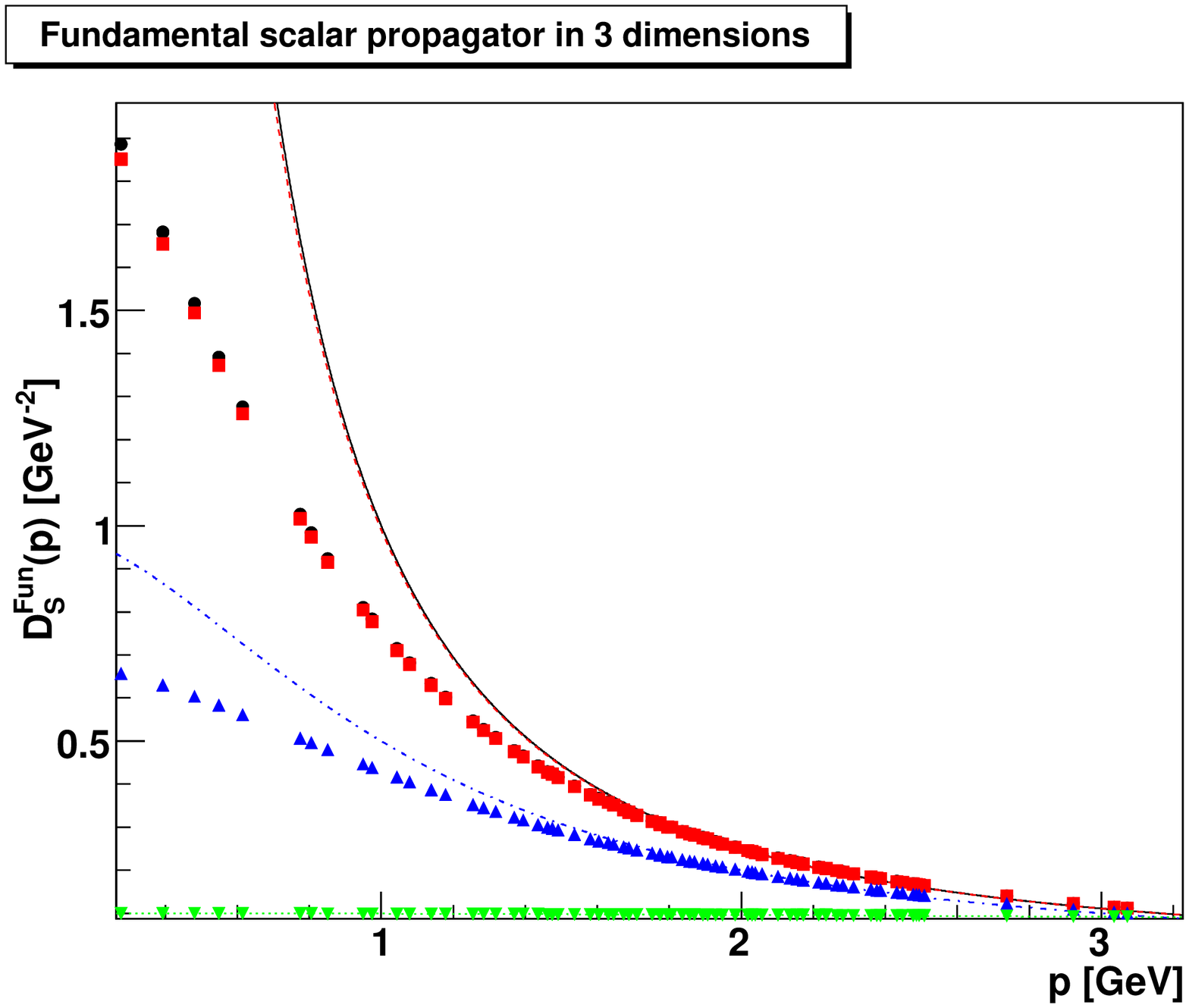}\includegraphics[width=0.5\linewidth]{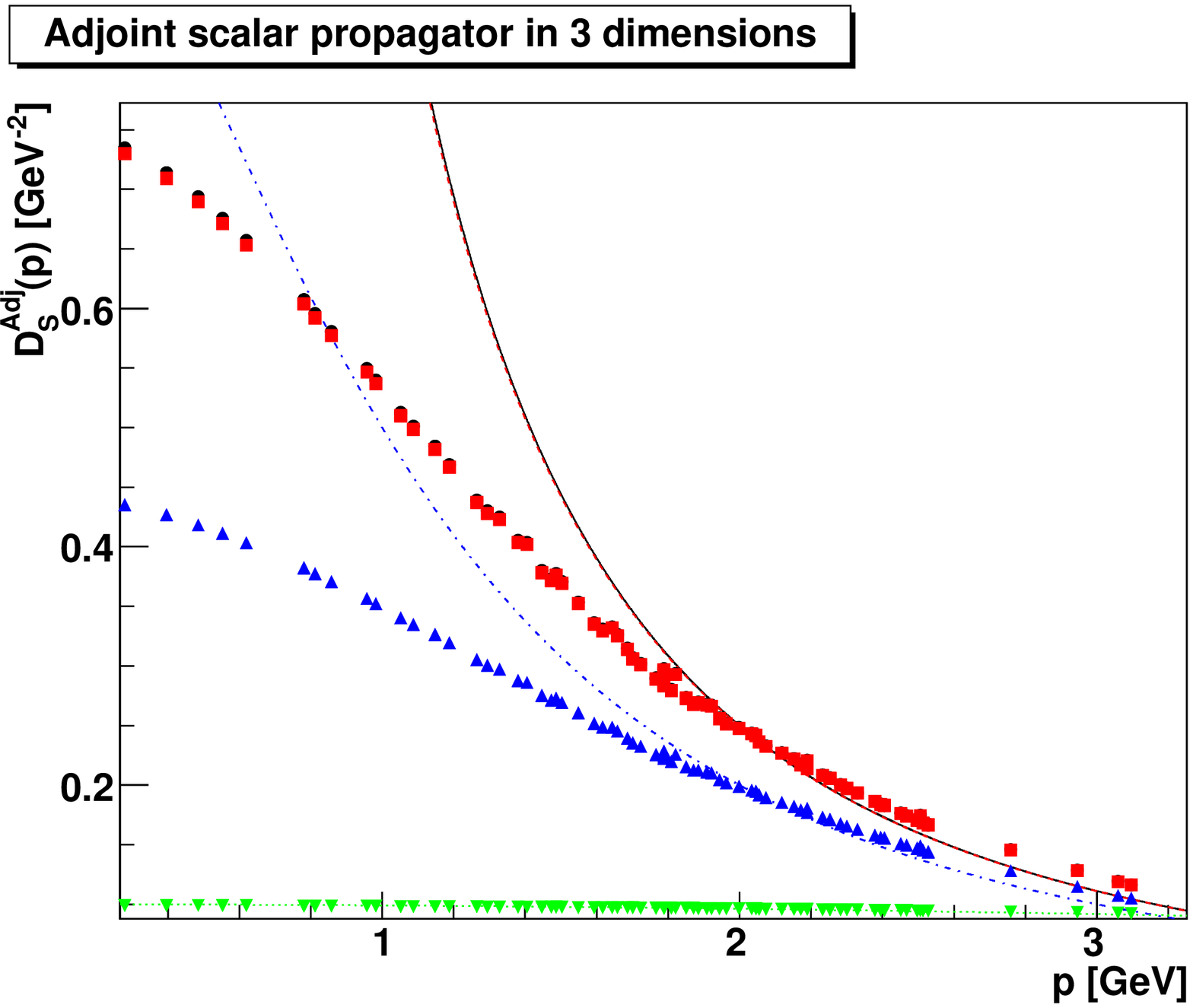}\\
\includegraphics[width=0.5\linewidth]{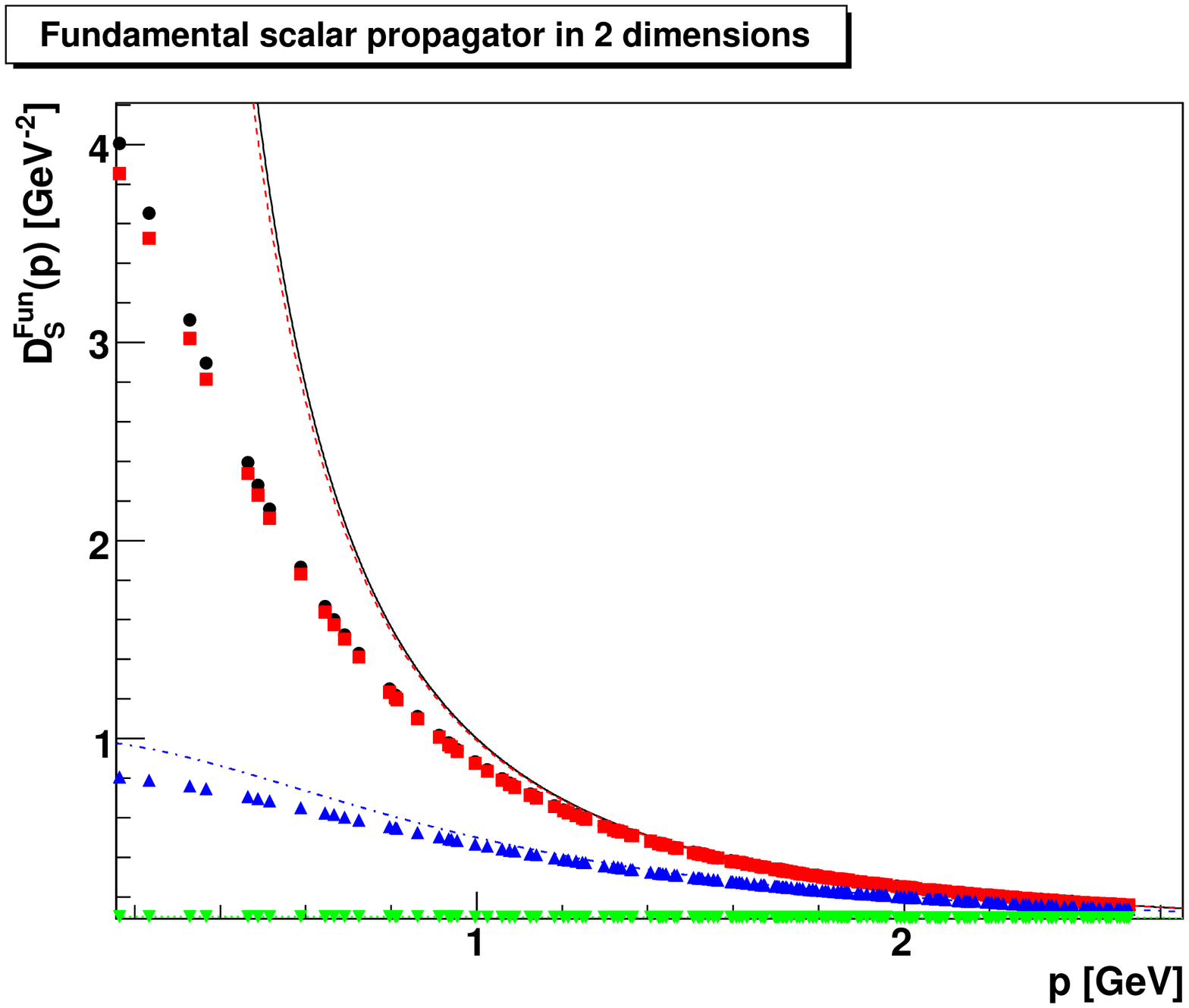}\includegraphics[width=0.5\linewidth]{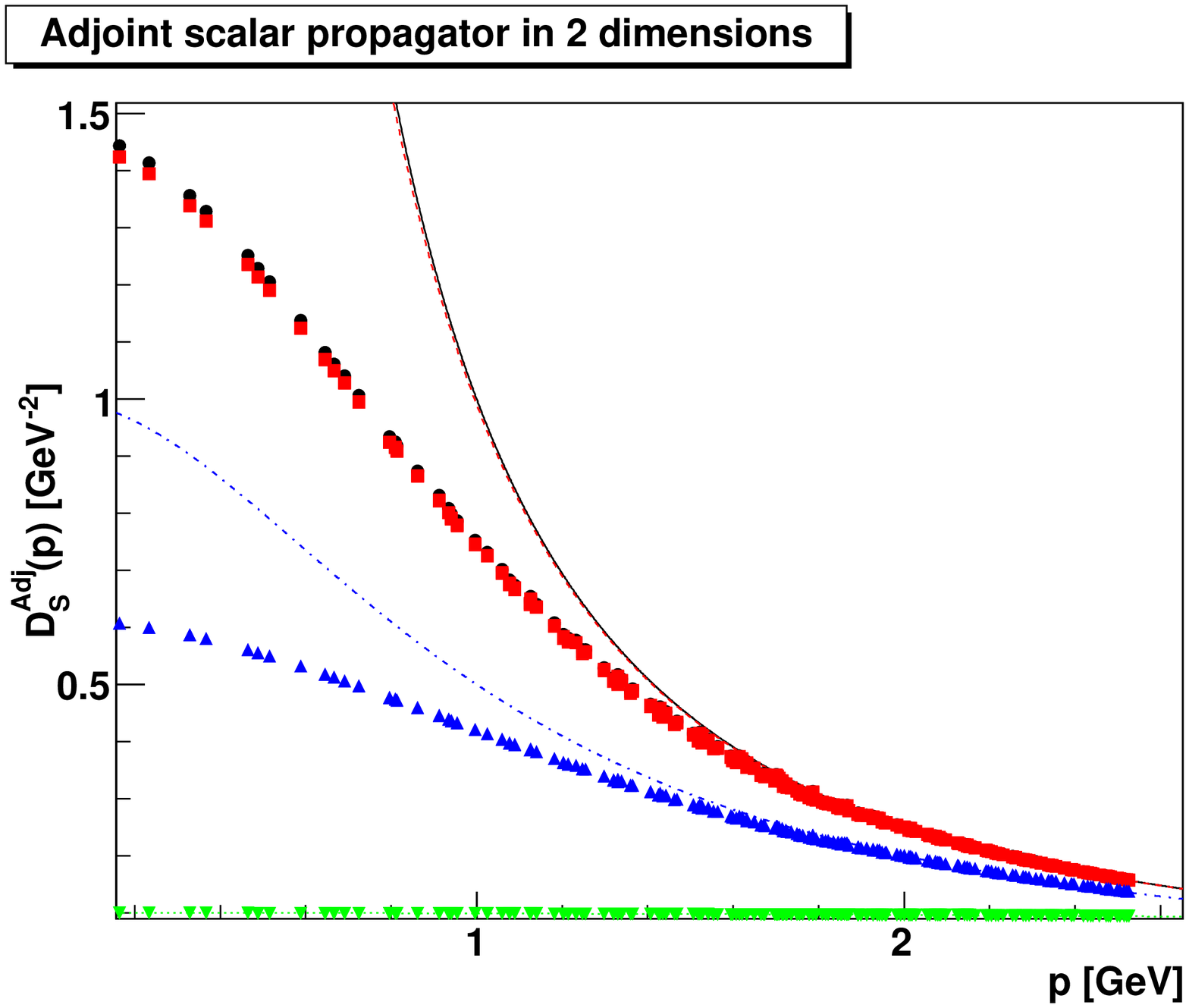}
\caption{\label{fig:prop}The propagators of the quenched scalars. The left panels show results for fundamental scalars and the right panels for adjoint scalars. The top panels are in four dimensions with lattice size $14^4$ at $a^{-1}=0.9$ GeV, the middle panels are in three dimensions with lattice size $20^3$ at $a^{-1}=0.9$ GeV and the bottom panels are in two dimensions with lattice size $34^2$ at $a^{-1}=0.9$ GeV. The results are always compared to the tree-level behavior. Statistical error bars are smaller than the symbol size.}
\end{figure}

The propagators for the four different masses for both representations and in two, three, and four dimensions are shown in figure \ref{fig:prop}. Note that the color-averaged propagators are shown. The off-diagonal propagators are zero within errors \cite{Maas:unpublished}, as is expected from the color structure of the corresponding Dyson-Schwinger equations \cite{Macher:2011ad,Macher:2010di}.

Immediately a number of interesting observations can be made. First, the lower the dimension, the more tree-level-like are the results. Secondly, at large momenta the adjoint propagator deviates stronger from the tree-level one than the fundamental one. This effect increases with the dimensionality. The origin of this effect may largely be due to perturbative effects. Especially the increase with dimensionality may be due to renormalization effects.

However, the deviations at large momenta are by far not as dramatic as the deviations at small momenta. The latter increase with decreasing tree-level mass. This is naively expected as for larger tree-level mass a particle should be less sensitive to infrared effects, and should decouple, at least perturbatively \cite{Collins:1984xc}. Given that even the largest volume still has a lowest momentum of more than 100 MeV, it is then consequently also not too surprising that the results for tree-level mass zero and 100 MeV do not differ significantly, which is already seen from the tree-level case.

The most interesting result is, however, that the propagators show an additional screening effect compared to tree-level, i.\ e.\ their screening mass $D(0)^{-1/2}$ appears to be larger than their renormalized mass. Of course, larger volumes and a more reliable extrapolation to zero momentum are needed to confirm this. In particular whether a non-zero screening mass exists for zero tree-level mass remains to be seen. Assuming that the current results faithfully extrapolate to zero momentum, it is a very interesting observation that the screening mass depends significantly on the bare mass. E.\ g., in the adjoint case for two dimensions the screening mass for 100 MeV tree-level mass could be as large as 800 MeV, while it is only as small as 1300 MeV for 1000 MeV tree-level mass. This is not observed for quarks \cite{Fischer:2006ub}. Furthermore, the screening mass in the adjoint case seems to be larger than in the fundamental case, e.\ g., 1.1 vs.\ 1.3 GeV in two dimensions for 1 GeV tree-level mass. This, on the other hand, is also observed in the comparison of adjoint and fundamental quarks \cite{Aguilar:2010ad}.

However, it should be kept in mind that in contrast to fermions there is no distinction between a mass function and a wave-function, just as for gluons. Therefore, the screening mass contains admixtures from both the wave-function renormalization and the mass-renormalization, and should even less than in the case for fermions be put into a connection with a possibly existing pole mass. Nonetheless, the screening mass, defined in the above way, is still a measure of how much long-distance modes are screened, when taken as a ratio with other quantities such that any renormalization-group dependency cancels. Here it is mainly an indication of how much the behavior of the propagator deviates from the tree-level behavior, which is imposed at the renormalization point, and thus the relevant comparison is made between the propagator at the renormalization point and zero momentum.

Summarizing, the quenched scalar propagator shows a significant departure from tree-level, and it appears that scalars are screened. However, a careful analysis  of finite-volume and cutoff effects will be necessary for any conclusive results on the continuum and infinite-volume limit \cite{Maas:unpublished}.

\subsection{Three-point vertices}

\begin{figure}
\includegraphics[width=0.5\linewidth]{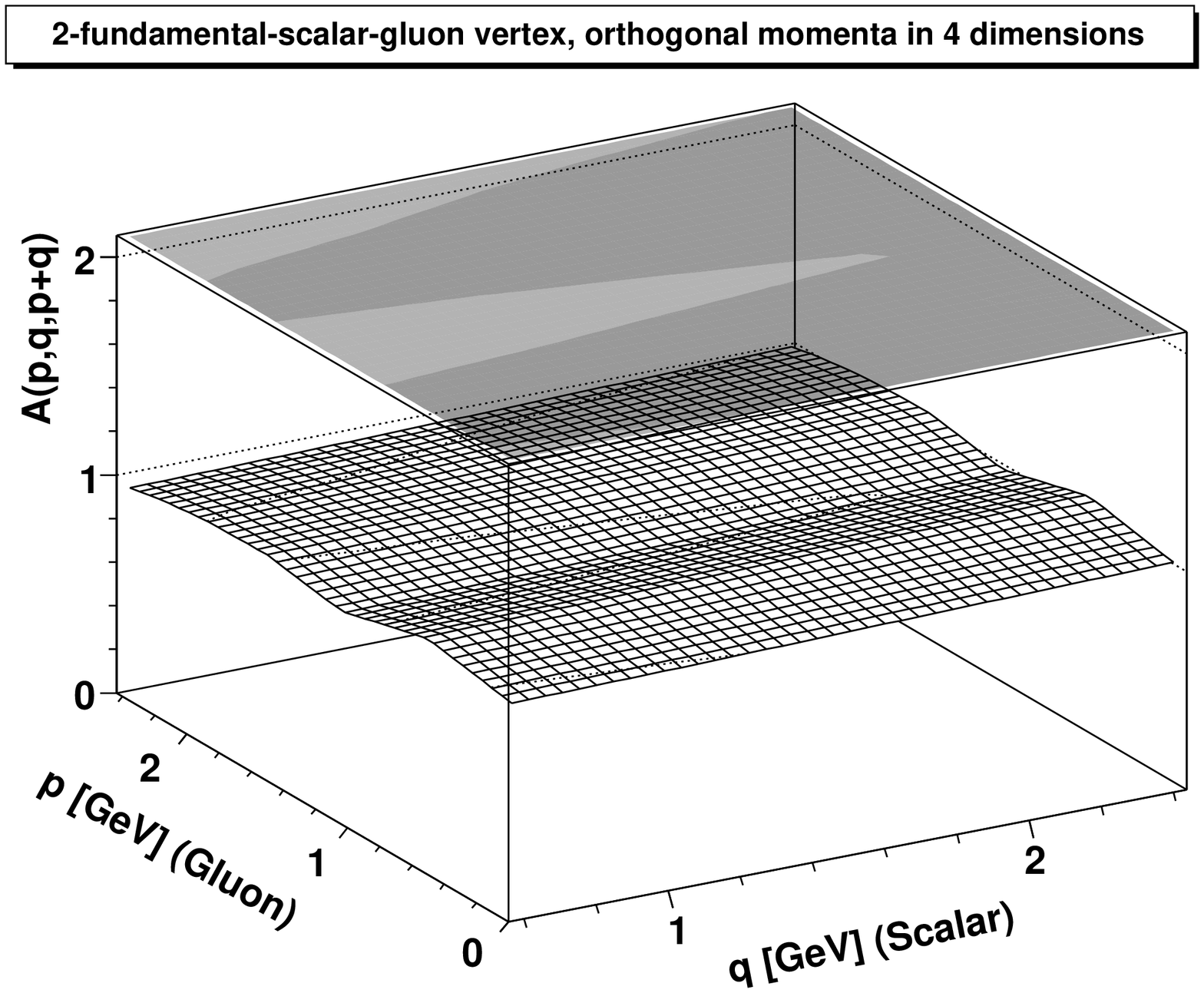}\includegraphics[width=0.5\linewidth]{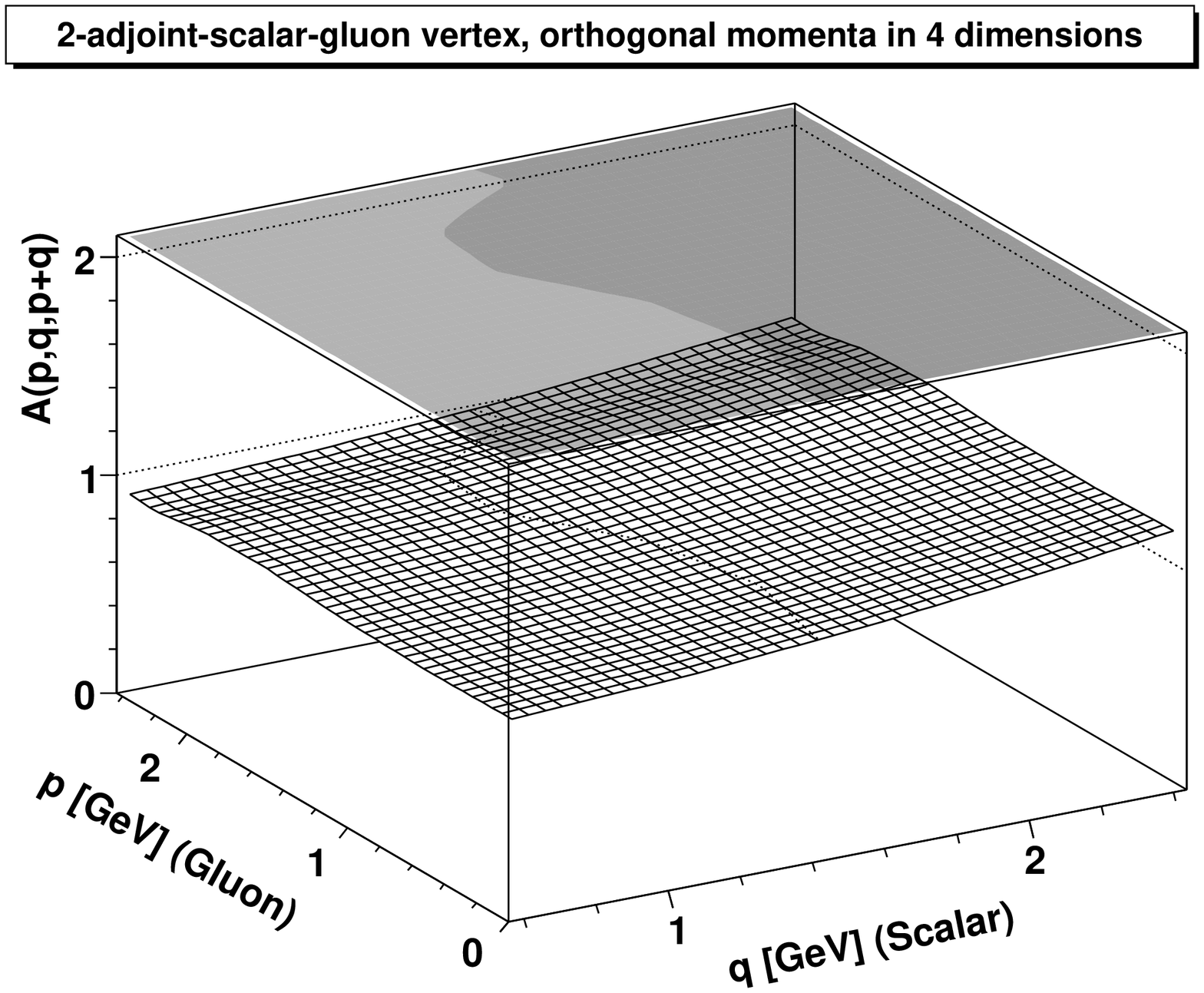}\\
\includegraphics[width=0.5\linewidth]{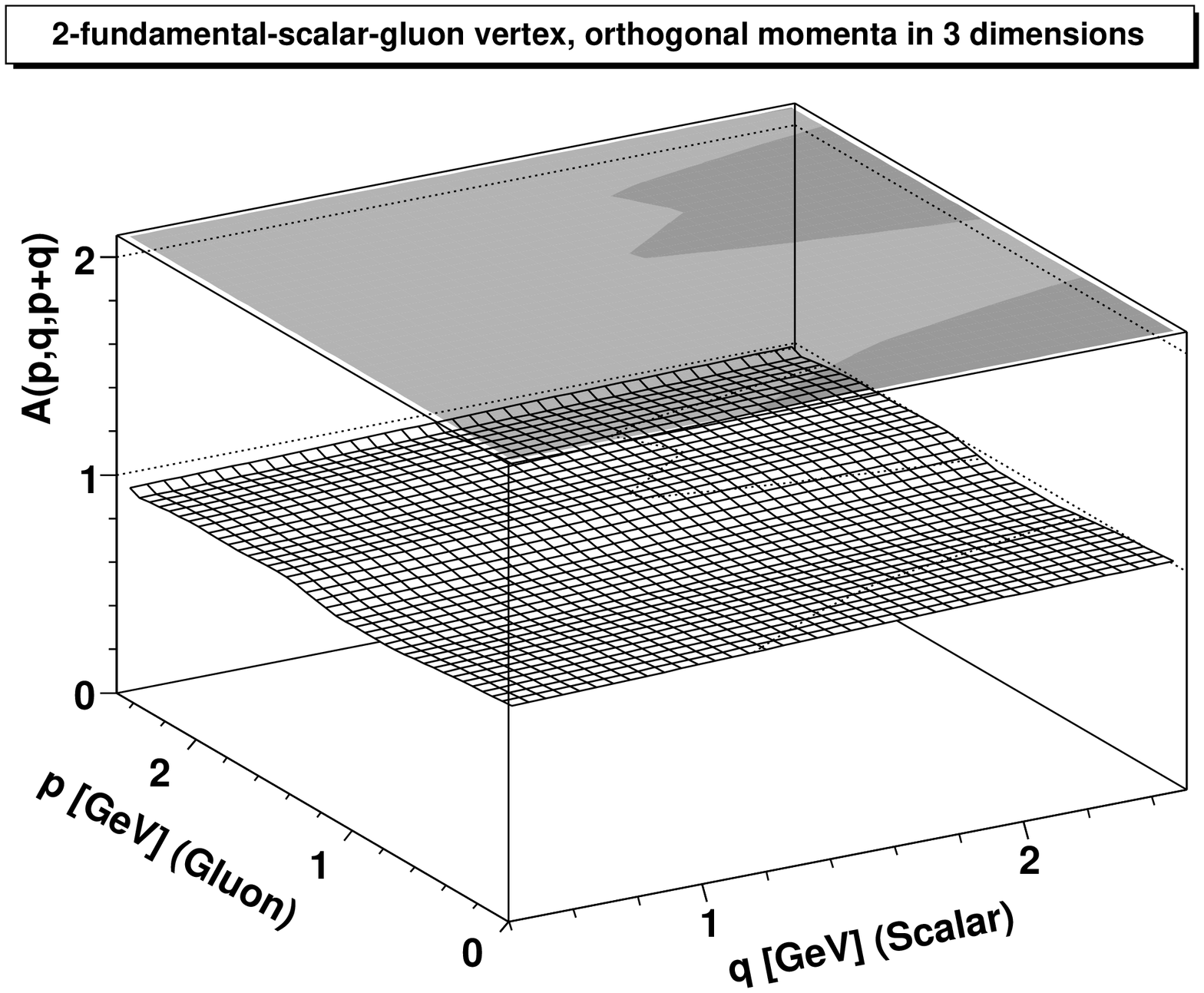}\includegraphics[width=0.5\linewidth]{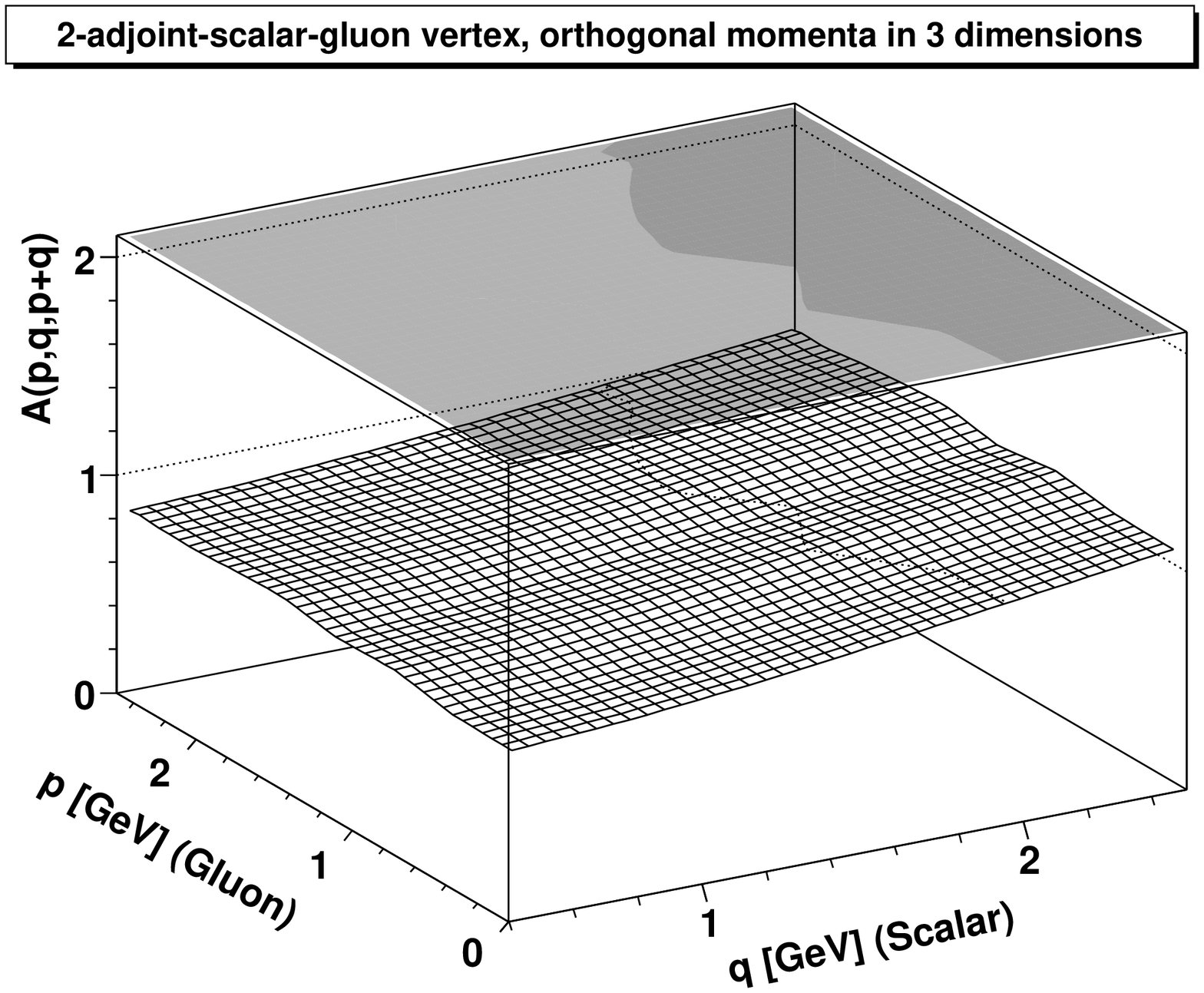}\\
\includegraphics[width=0.5\linewidth]{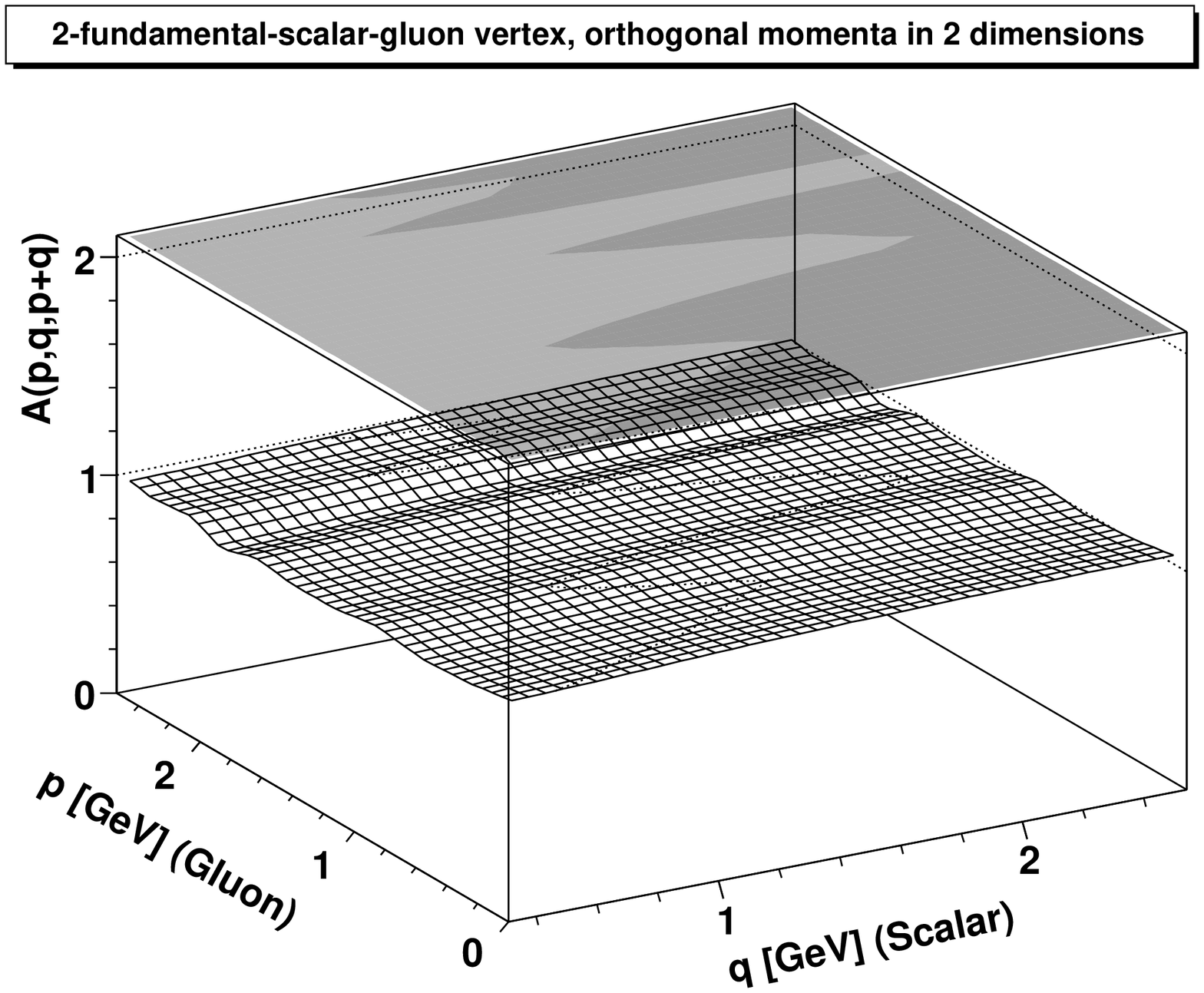}\includegraphics[width=0.5\linewidth]{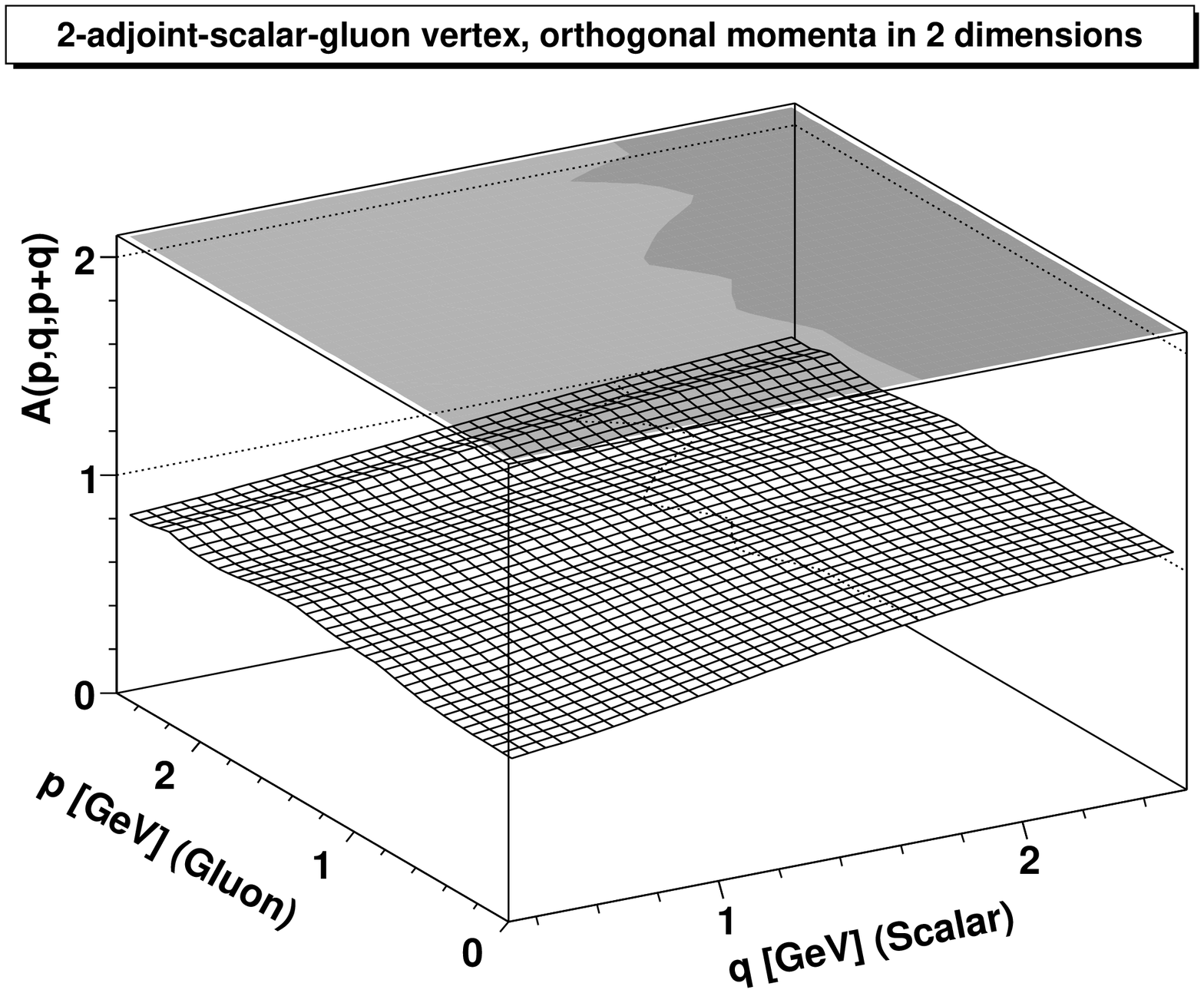}
\caption{\label{fig:vert}The single transverse tensor structure of the two-scalar-gluon vertex for the momentum configuration $pq=0$. The left panels show results for the fundamental scalar and the right panels for adjoint scalars. The top panels are in four dimensions with lattice size $14^4$ at $a^{-1}=1.3$ GeV, the middle panels are in three dimensions with lattice size $20^3$ at $a^{-1}=1.3$ GeV and the bottom panels are in two dimensions with lattice size $26^2$ at $a^{-1}=1.3$ GeV. The results shown are for the case of 100 MeV tree-level mass. Heavier masses show a behavior closer to tree-level, while the zero tree-level mass case is almost indistinguishable for these lattice parameters \cite{Maas:unpublished}.}
\end{figure}

The results for the two-scalar-gluon vertex for the momentum configuration $pq=0$ are shown in figure \ref{fig:vert}. The results for the symmetric case $p^2=q^2=(p+q)^2$ do not show a strongly different behavior for the current lattice settings \cite{Maas:unpublished}.

Investigating the vertex, it shows a behavior very close to tree-level, even closer than the ghost-gluon vertex \cite{Cucchieri:2008qm,Maas:2007uv}. If at all, there is only a slight suppression for small scalar momenta, though this effect is partly influenced by cutoff effects \cite{Maas:unpublished}. This effect is stronger for the adjoint case than for the fundamental case, and there also appears to be some dependency on the dimensionality, but the latter is rather weak. Furthermore, there is almost no dependency on the gluon momentum. In particular, there is no sign of possible collinear singularities \cite{Fister:2010yw}, which is not unexpected \cite{Fischer:2009tn}. However, this statement should be taken with great care, as the lowest accessible momenta are still very large, and strictly speaking the current gauge is not expected \cite{Maas:2009se} to show the behavior obtained in \cite{Fister:2010yw}.

This is a rather disappointing result, given that one could have hoped that infrared divergencies in the two-scalar-gluon vertex may be relevant for the Wilson-line behavior, or that at least the distinction of fundamental and adjoint matter should be imprinted clearly on this vertex. On the other hand, such a slight deviation from tree-level is of course very useful for the design of truncations as used in functional methods \cite{Maas:2004se}.

\section{Comparison to unquenched scalars}\label{sec:unquenched}

In summary, in the quenched case there are significant deviations from the tree-level behavior for the propagators, and a mild deviation for the two-scalar-gluon vertex from the tree-level vertex. To fully investigate scalar matter, unquenching is necessary. With dynamical scalar matter, no qualitative difference exists anymore for the Wilson line between fundamental and adjoint matter, and string breaking occurs in both cases by means of screening through bound-state formation. In the case of fundamental quarks this change leaves only a minor imprint on the Yang-Mills sector and the quark sector, at least for the propagators \cite{Fischer:2006ub,Kamleh:2007ud}.

\begin{figure}
\includegraphics[width=0.5\linewidth]{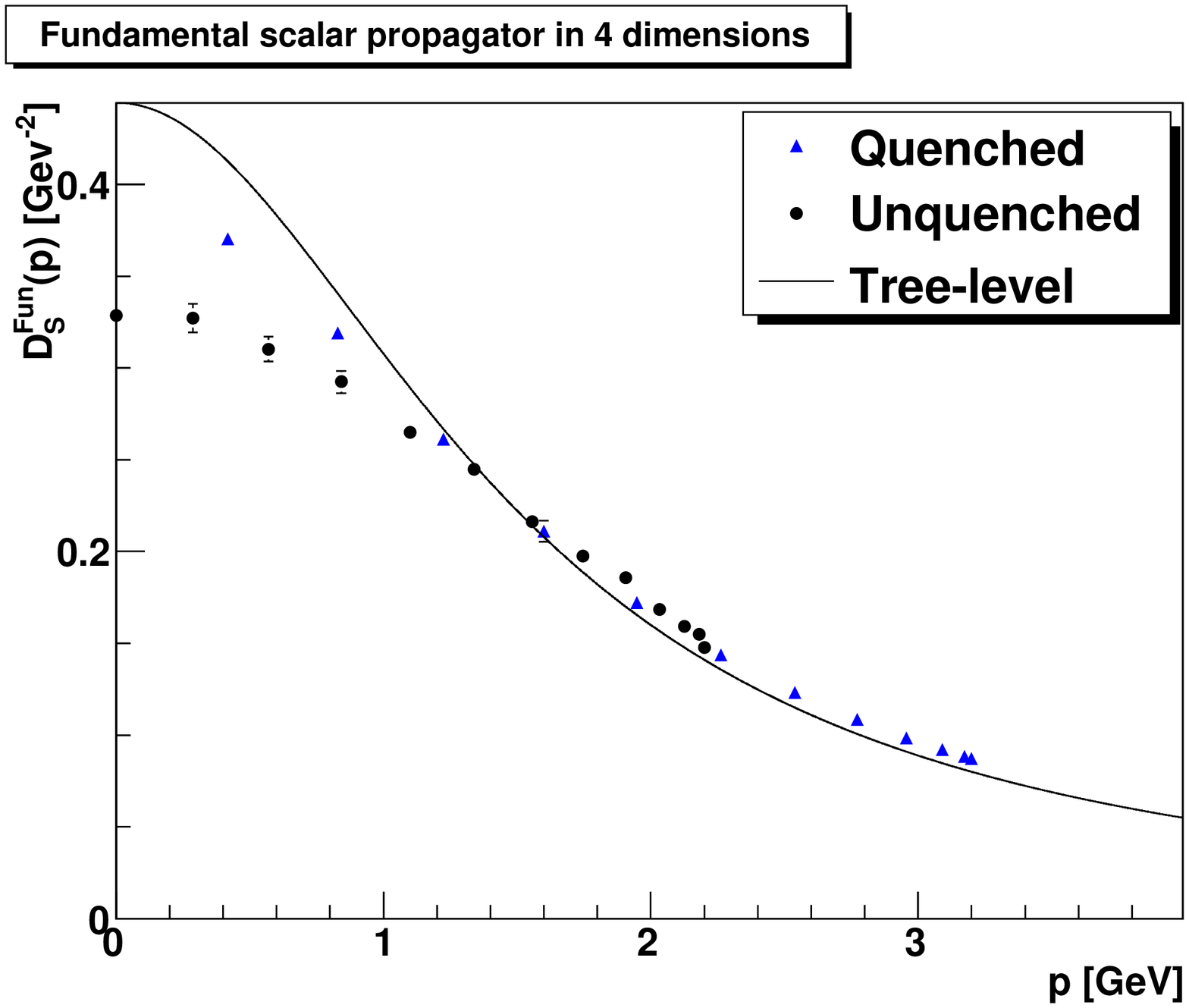}\includegraphics[width=0.5\linewidth]{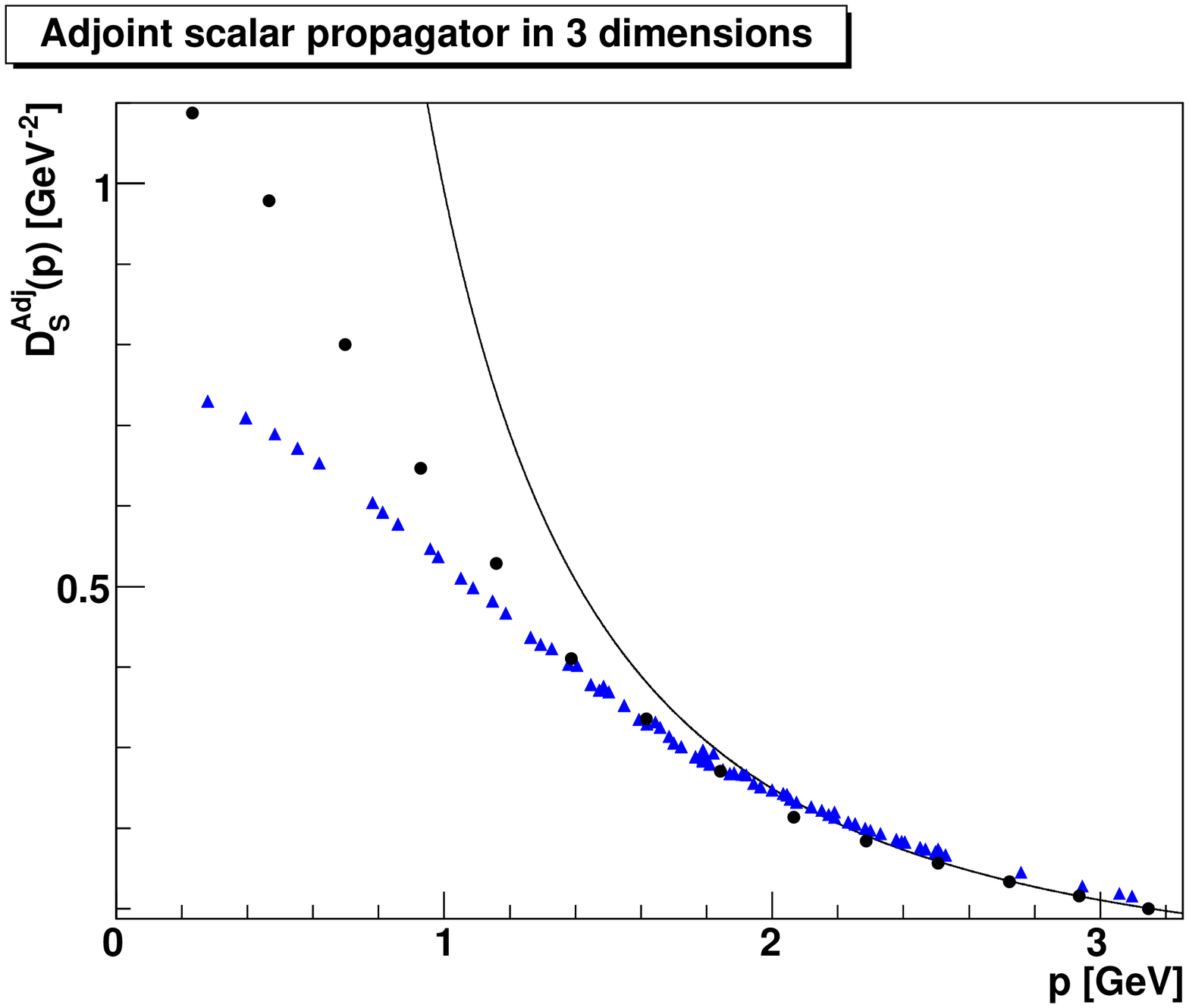}
 \caption{\label{fig:comp}In the left panel the quenched and unquenched fundamental scalar propagator in four dimensions are compared, the unquenched results being from \cite{Maas:2010nc}. In the right panel the same is done for the adjoint scalar propagator in three dimensions with the unquenched results from \cite{Cucchieri:2001tw}.}
\end{figure}

Currently, calculation exist for the unquenched case for the fundamental propagator in four dimensions \cite{Maas:2010nc} and the adjoint propagator in three dimensions \cite{Cucchieri:2001tw}. It is of course very complicated to assess the correct translation of the corresponding scale, or even set to the corresponding scale in the first place \cite{Maas:2010nc}. Therefore, at best a qualitative comparison can be made, using here a bare mass of 1.5 GeV in four dimensions \cite{Maas:2010nc} and 100 MeV in three dimensions. This is shown in figure \ref{fig:comp}. In both cases the comparison is done in what is usually called the confinement phase, though this name has to be taken with care \cite{Caudy:2007sf}, in particular for the fundamental case. The change in the Higgs phase does not appear to be too large for the scalar propagator \cite{Maas:2010nc}, though the influence on the Yang-Mills sector is substantial \cite{Maas:2010nc} and has to be investigated further.

Returning to the comparison in figure \ref{fig:comp}, it is visible that, as far as this crude comparison can be taken, there is no significant qualitative difference between the quenched and unquenched case, though the involved scales may be significantly different. This confirms the picture from the fermion case that unquenching may have only rather small effects on the matter propagator, at least when the overall properties of the systems are similar enough to the quenched case, in particular concerning asymptotic freedom and the state of global symmetries. On the other hand that implies once more that it is unlikely that the propagators alone are sufficient to learn about the difference of fundamental and adjoint matter. Studying at least the three-point vertices will be an important next step.

\section{Summary}\label{sec:sum}

Summarizing, it is possible and feasible to determine at least the propagators and three-point vertices for quenched scalar matter in the fundamental and adjoint representation. Furthermore, the results show, at least for the lattice settings studied, no obvious differences between both cases. If this were confirmed on larger volumes and finer cutoffs and eventually in the continuum and infinite-volume limit, this would imply that higher-order correlation functions may be necessary to understand the origin of the difference between adjoint and fundamental matter. However, even so innocent looking propagators as those shown here can have rather interesting analytical structures \cite{Maas:2004se,Maas:2005rf}, and thus a detailed analysis with improved lattice settings is an important step \cite{Maas:unpublished}.

These investigations provide interesting starting points for calculations using functional methods \cite{Maas:2004se}. If indeed there is no extreme behavior in the two-scalar-gluon vertex, this would lead to a very simple truncation for such calculations. Since the tensor structure of scalars is furthermore simple, this may be an important testbed to develop control over truncation artifacts and for the improvement of numerical methods.

\bibliographystyle{bibstyle}
\bibliography{bib}

\begin{thebibliography}{10}

\bibitem{Fischer:2006ub}
C.~S. Fischer,
\newblock J. Phys. {\bf G32}, R253 (2006), hep-ph/0605173.

\bibitem{Maas:2010gi}
A.~Maas,
\newblock (2010), 1011.5409.

\bibitem{Alkofer:2008tt}
R.~Alkofer, C.~S. Fischer, F.~J. Llanes-Estrada, and K.~Schwenzer,
\newblock Annals Phys. {\bf 324}, 106 (2009), 0804.3042.

\bibitem{Kamleh:2007ud}
W.~Kamleh, P.~O. Bowman, D.~B. Leinweber, A.~G. Williams, and J.~Zhang,
\newblock Phys. Rev. {\bf D76}, 094501 (2007), 0705.4129.

\bibitem{Kizilersu:2006et}
A.~Kizilersu, D.~B. Leinweber, J.-I. Skullerud, and A.~G. Williams,
\newblock Eur. Phys. J. {\bf C50}, 871 (2007), hep-lat/0610078.

\bibitem{Aguilar:2010ad}
A.~C. Aguilar and J.~Papavassiliou,
\newblock Phys. Rev. {\bf D83}, 014013 (2011), 1010.5815.

\bibitem{Zwanziger:2010iz}
D.~Zwanziger,
\newblock Phys. Rev. {\bf D81}, 125027 (2010), 1003.1080.

\bibitem{Fister:2010yw}
L.~Fister, R.~Alkofer, and K.~Schwenzer,
\newblock Phys. Lett. {\bf B688}, 237 (2010), 1003.1668.

\bibitem{Alkofer:2010tq}
R.~Alkofer, L.~Fister, A.~Maas, and V.~Macher,
\newblock (2010), 1011.5831.

\bibitem{Maas:2010nc}
A.~Maas,
\newblock Eur. Phys. J. C in print  (2010), 1007.0729.

\bibitem{Caudy:2007sf}
W.~Caudy and J.~Greensite,
\newblock Phys. Rev. {\bf D78}, 025018 (2008), 0712.0999.

\bibitem{Philipsen:1998de}
O.~Philipsen and H.~Wittig,
\newblock Phys. Rev. Lett. {\bf 81}, 4056 (1998), hep-lat/9807020.

\bibitem{Callaway:1988ya}
D.~J.~E. Callaway,
\newblock Phys. Rept. {\bf 167}, 241 (1988).

\bibitem{Greensite:2003bk}
J.~Greensite,
\newblock Prog. Part. Nucl. Phys. {\bf 51}, 1 (2003), hep-lat/0301023.

\bibitem{Alkofer:2003jj}
R.~Alkofer, W.~Detmold, C.~S. Fischer, and P.~Maris,
\newblock Phys. Rev. {\bf D70}, 014014 (2004), hep-ph/0309077.

\bibitem{Maas:2004se}
A.~Maas, J.~Wambach, B.~Gr\"uter, and R.~Alkofer,
\newblock Eur. Phys. J. {\bf C37}, 335 (2004), hep-ph/0408074.

\bibitem{Maas:2005rf}
A.~Maas,
\newblock {\em {The high-temperature phase of Yang-Mills theory in Landau
  gauge}},
\newblock PhD thesis, Darmstadt University of Technology, 2004, hep-ph/0501150.

\bibitem{Cucchieri:2001tw}
A.~Cucchieri, F.~Karsch, and P.~Petreczky,
\newblock Phys. Rev. {\bf D64}, 036001 (2001), hep-lat/0103009.

\bibitem{Gattringer:2010zz}
C.~Gattringer and C.~B. Lang,
\newblock {\em Quantum chromodynamics on the lattice} (Lect. Notes Phys.,
  2010).

\bibitem{Cucchieri:2006tf}
A.~Cucchieri, A.~Maas, and T.~Mendes,
\newblock Phys. Rev. {\bf D74}, 014503 (2006), hep-lat/0605011.

\bibitem{Maas:unpublished}
A.~Maas,
\newblock unpublished.

\bibitem{Cucchieri:2004sq}
A.~Cucchieri, T.~Mendes, and A.~Mihara,
\newblock JHEP {\bf 12}, 012 (2004), hep-lat/0408034.

\bibitem{Fischer:2008uz}
C.~S. Fischer, A.~Maas, and J.~M. Pawlowski,
\newblock Annals Phys. {\bf 324}, 2408 (2009), 0810.1987.

\bibitem{Maas:2009se}
A.~Maas,
\newblock Phys. Lett. {\bf B689}, 107 (2010), 0907.5185.

\bibitem{Maas:2010wb}
A.~Maas,
\newblock (2010), arXiv:1010.5718.

\bibitem{Cucchieri:2007rg}
A.~Cucchieri and T.~Mendes,
\newblock Phys. Rev. Lett. {\bf 100}, 241601 (2008), 0712.3517.

\bibitem{Cucchieri:2008fc}
A.~Cucchieri and T.~Mendes,
\newblock Phys. Rev. {\bf D78}, 094503 (2008), 0804.2371.

\bibitem{Bogolubsky:2009dc}
I.~L. Bogolubsky, E.~M. Ilgenfritz, M.~M{\"u}ller-Preussker, and A.~Sternbeck,
\newblock Phys. Lett. {\bf B676}, 69 (2009), 0901.0736.

\bibitem{Binosi:2009qm}
D.~Binosi and J.~Papavassiliou,
\newblock Phys. Rept. {\bf 479}, 1 (2009), 0909.2536.

\bibitem{Bornyakov:2008yx}
V.~G. Bornyakov, V.~K. Mitrjushkin, and M.~M\"uller-Preussker,
\newblock Phys. Rev. {\bf D79}, 074504 (2009), 0812.2761.

\bibitem{Boucaud:2008ji}
P.~Boucaud {\em et~al.},
\newblock JHEP {\bf 06}, 012 (2008), 0801.2721.

\bibitem{Dudal:2008sp}
D.~Dudal, J.~A. Gracey, S.~P. Sorella, N.~Vandersickel, and H.~Verschelde,
\newblock Phys. Rev. {\bf D78}, 065047 (2008), 0806.4348.

\bibitem{Maas:2007uv}
A.~Maas,
\newblock Phys. Rev. {\bf D75}, 116004 (2007), 0704.0722.

\bibitem{Macher:2011ad}
V.~Macher, A.~Maas, and R.~Alkofer,
\newblock unpublished.

\bibitem{Macher:2010di}
V.~Macher,
\newblock Adjoint and scalar fields coupled to landau-gauge yang-mills theory,
\newblock Master's thesis, Karl-Franzens-University Graz, 2010.

\bibitem{Collins:1984xc}
J.~C. Collins,
\newblock {\em {Renormalization: An introduction to renormalization, the
  renormalization group, and the operator product expansion}} (Cambrdige
  University Press, Cambridge, 1984).

\bibitem{Cucchieri:2008qm}
A.~Cucchieri, A.~Maas, and T.~Mendes,
\newblock Phys. Rev. {\bf D77}, 094510 (2008), 0803.1798.

\bibitem{Fischer:2009tn}
C.~S. Fischer and J.~M. Pawlowski,
\newblock Phys. Rev. {\bf D80}, 025023 (2009), 0903.2193.

\end{thebibliography}

\end{document}